\documentclass{ws-procs9x6-cpt19}

\begin{document}

\newcommand{\refeq}[1]{(\ref{#1})}
\def\etal {{\it et al.}}
%any other macros go here 

\title{Lorentz Violation in the Matter--Gravity Sector
}

\author{Zonghao Li}

\address{Physics Department, Indiana University,
Bloomington, IN 47405, USA}

\begin{abstract}
We construct the general Lorentz-violating effective field theory in curved spacetime
and the corresponding nonrelativistic Hamiltonian in the Earth's gravitational field.
Applying this general framework to three types of experiments,
free-dropping, interferometer, and bound-state experiments,
we extract first constraints on certain new coefficients 
in the matter--gravity sector.
\end{abstract}

\bodymatter

\section*{}

Lorentz symmetry is a fundamental symmetry 
in both General Relativity (GR) and the Standard Model (SM); 
it deserves to be precisely tested in experiments.
Moreover,
Lorentz violation has been a popular candidate in recent years
as a low-energy remnant of the unification 
of GR and the SM.
To study Lorentz violation,
D.\ Colladay and V.A.\ Kosteleck\'y 
developed a comprehensive framework,
the Standard-Model Extension (SME),
in the context of effective field theory.\cite{ck97,ak04}
The SME contains GR and the conventional SM coupled to GR,
and it adds all possible Lorentz-violating modifications
at the Lagrangian level.
We have constructed the general Lorentz-violating terms in flat spacetime.\cite{kl19}
Expanding this method,
we further built the general Lorentz-violating terms in curved spacetime.\cite{kl20, kl21}

We first sketch the procedure of building the Lagrange density
and then investigate its applications to experiments.
We focus on the experiments in the matter--gravity sector,
which has been a fruitful area in the search for Lorentz-violating signals.\cite{kt11}
Using the general framework we built,
we can study some new effects,
including those from nonminimal coefficients and spin--gravity couplings.
The present contribution to the CPT'19 proceedings 
is based on results in Ref.\ \refcite{kl20} and Ref.\ \refcite{kl21}.

In flat spacetime,
the general Lorentz-violating effective field theory
is built from gauge-covariant operators 
to preserve gauge invariance.\cite{kl19}
In curved spacetime,
this idea is expanded to gauge-covariant spacetime-tensor operators
to incorporate observer diffeomorphism invariance.
Coupling those operators with coefficients for Lorentz violation,
we construct the full SME in curved spacetime.

To make applications to experiments easier, 
we transfer the full Lagrange density into its nonrelativistic Hamiltonian.
We first turn the full Lagrange density into its linearized Lagrange density
by taking the linearized limit in a weak gravity field.
Then,
we extract the relativistic Hamiltonian
by solving the equation of motion.
Finally,
we get the nonrelativistic Hamiltonian by the Foldy--Wouthuysen transformation,\cite{fw50}
which is a systematic procedure to extract nonrelativistic Hamiltonians
from relativistic Hamiltonians for fermions.

The nonrelativistic Hamiltonian for a spin-1/2 fermion
in the Earth's gravitational field
can be expressed as:
\begin{equation}
\label{eq:ph}
H=H_0+H_\phi+H_g+H_{\sigma\phi}+H_{\sigma g},
\end{equation}
where $H_0$ is the conventional term,
and the remaining terms are corrections from the SME
and depend on the coefficients for Lorentz violation.
These corrections contain couplings between
the gravitational field 
and the position, momentum, and spin of the particle.
Terms in $H_\phi$ and $H_g$
are independent of the spin,
but those in $H_{\sigma\phi}$ and $H_{\sigma g}$
depend on the spin,
which permits us to study spin--gravity couplings in the SME framework
for the first time.

Three types of experiments,
free-dropping, interferometer, and bound-state experiments,
are analyzed as examples of the applications of the general framework 
in the matter--gravity sector.

Free-dropping experiments compared the accelerations
of freely falling atoms with different inner structures.
We focus on free drops of atoms with different spin polarizations
to test spin--gravity couplings.
An experiment in Italy dropped unpolarized $^{87}$Sr atoms
and tested the broadening of the accelerations
due to different spin polarizations.\cite{mt14}
Another experiment in China dropped $^{87}$Rb atoms
with different spin polarizations
and measured the difference between the accelerations of the atoms.\cite{xd16}
The relative discrepancies of accelerations are bounded around $10^{-7}$
in these experiments.
These are transformed into bounds on certain coefficients 
for Lorentz violation and spin--gravity couplings.\cite{kl20}
Notice that different experiments constrain different coefficients for Lorentz violation;
more experiments,
such as dropping hydrogen and antihydrogen,\cite{antihydrogen}
are needed for full coverage of those coefficients.

Interferometer experiments measured
gravity-induced quantum phase shifts.
Using the classic COW experiment
performed by R.\ Colella, A.W.\ Overhauser, and S.A.\ Werner
in 1975,\cite{cow75}
we get bounds on certain spin-independent coefficients.\cite{kl20}
More recent interferometer experiments\cite{vh14}
using magnetic fields to split neutron beams
are expected to improve the result 
and constrain some spin-dependent coefficients. 

Bound-state experiments measured 
the energies of the bound states of neutrons in the Earth's gravitational field.
The original experiment measured the critical heights,\cite{vn02}
which are related to the energies by $mgz=E$.
The improved version measured the transition frequencies
between different energy states.\cite{gc18}
A sensitivity of $10^{-2}$ was attained
and is used to constrain certain coefficients 
for Lorentz violation and spin--gravity couplings.\cite{kl20}

In summary,
fruitful results have been extracted from these three types of experiments.
More experiments are needed for full coverage 
of the coefficients for Lorentz violation and spin--gravity couplings.

\section*{Acknowledgments}
This work was supported in part by the US Department of Energy
and by the Indiana University Center for Spacetime Symmetries.

\end{document}